\documentclass[conference]{IEEEtran}

\usepackage{cite}
\usepackage{amsmath,amssymb,amsfonts}
\usepackage{algorithmic}
\usepackage{graphicx}
\usepackage{textcomp}
\graphicspath{ {./images/} }

\author{\IEEEauthorblockN{Lakshmi Sirisha Revadi}
\IEEEauthorblockA{\textit{School of Information Technology} \\
\textit{Deakin University}\\
Melbourne, Australia \\
revadi.sirisha@gmail.com}
\and
\IEEEauthorblockN{Xi Zheng}
\IEEEauthorblockA{\textit{Department of Computing} \\
\textit{Macquarie University}\\
Sydney, Australia \\
james.zheng@mq.edu.au}
\and
\IEEEauthorblockN{Yupeng Jiang}
\IEEEauthorblockA{\textit{Department of Computing} \\
\textit{Macquarie University}\\
Sydney, Australia \\
yupeng.jiang@hdr.mq.edu.au}
}
\title{Investigating Key User Experiencing Engineering Aspects in Software-as-a-Service Service Delivery Model}

\begin{document}
\maketitle

\begin{abstract}
Software as a Service (SaaS) is well established as an effective model for the development, deployment and customization of software. As it continues to gain more momentum in the IT industry, many user experience challenges and issues are being reported by the experts and end users.
\end{abstract}

\section{Introduction}
In cloud computing, software as a service (SaaS) is one of the delivery models where any software is consumed and shared among various different tenants of the concerned service provider. For a SaaS application, the measure of the user experience is growing with end user goals and expanding changes \cite{kang2011design}. User experience is one of the major issue confronted by this model. Users are anticipating to have the capacity to utilize cloud based applications with various devices. They expect proficient collaboration flow and comprehensive inter-device user experience (UX). The most widely recognized user interfaces (UI) to cloud services contain web applications accessed through browsers and mobile applications. When considering mobile applications, the UI should be adjusted as per the screen size of the device, which would be a significant task in maintaining consistency of UX amongst various platforms and different devices \cite{Lamprecht2016}. The major problem in UI design is that not every UI design is easily accessible by the end users. As a result, various problems in user experience would arise like usability and design issue \cite{Lamprecht2016}.

\section{Existing Solutions}
In this section, we will walk through four techniques to deal with general UX challenges, which are human centered design technique (Section \ref{exsol:hc}), goal directed design method (Section \ref{exsol:gd}), advanced responsive design technique (Section \ref{exsol:rd}), and user experience design technique (Section \ref{exsol:ux}).

\subsection{Human Centered Design Technique}\label{exsol:hc}
Human-Centered design was developed to overcome the design issues by highlighting the needs and abilities of the end users, which usability and understandability of the interfaces has been indeed improved. Other than these advances, user experience complexity is still existing \cite{Elmansy2015}.

\subsubsection{Existing Solution}
\begin{figure}
\centering
\includegraphics[width=0.5\textwidth]{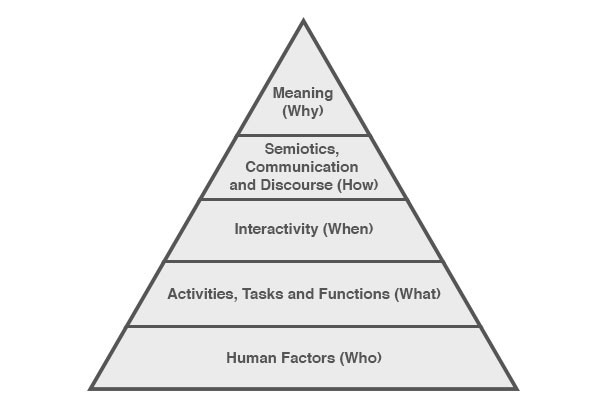}
\caption{The human centered design pyramid for interactive systems \cite{Giacomin2014}}
\label{fig:human_centered}
\end{figure}

According to \cite{Giacomin2014}, the Human-Centered design technique is very useful as it focuses on human factors, usability techniques, and iterative process to create interactive systems which can be more usable by users. As shown in Fig. \ref{fig:human_centered}, in this human centered design development, the design process has clear knowledge and understanding about the end users, tasks, and environments (as covered by these five dimensions --- Who, What, When, How, and Why) \cite{Elmansy2015}.

ISO (9241-210) defined six characteristics of the human centered design process.
\begin{itemize}
\item Clear knowledge and understanding about the end users, tasks and environments.
\item User-centered evaluation driven design.
\item Iterative design process.
\item Including the user in the design and considering overall user experience.
\end{itemize}

\subsubsection{Advantage}
There are many models with specific stages in the human centered design development process. These stages rely on the knowledge gained during the research and planning stages of good user experience. In these stages, a number of tools like persona, Scenario and Use cases can be utilized to evaluate the user behavior. However, as per the market the results are rewarding and successful in overcoming challenges \cite{Boy2017}.

\subsubsection{Limitation}
The major focus upon the users diminishes support for the implementations. Secondly, an excessive amount of consideration regarding the requirements of the users can be added complexity in the design. If something is tailored as per particular likes, skills, dislikes and needs, the less likely it will be appropriate for others \cite{Elmansy2015}.

\subsection{Goal Directed Design Method}\label{exsol:gd}
Goal directed design is a user-centered methodology where different users express a desire for different things \cite{Dubberly2001}. According to \cite{jelumalai2015}, this study approached users to browse a specific site, where only 29 percent of the participants were successful in their anticipated purchase. In the above scenario, 71 percent of the users were unable to locate the actual products they intended to buy due to lack of clarity in interaction and visual appeal.

\subsubsection{Existing Solution}
Goal directed design technique makes it possible for the users to research and try SaaS application before they are familiar with what it can do or how it is going to work. In designing good UI, it is always important to consider employee experience because it is where the first user experience begins \cite{Giurata2014}.

Flexibility of SaaS applications demonstrates how SaaS is not just an option for IT department but a very powerful instrument geared for the successful implementation of business strategy to attain great user experience \cite{Hassenzahl2006}.

\begin{figure}
\centering
\includegraphics[width=0.45\textwidth]{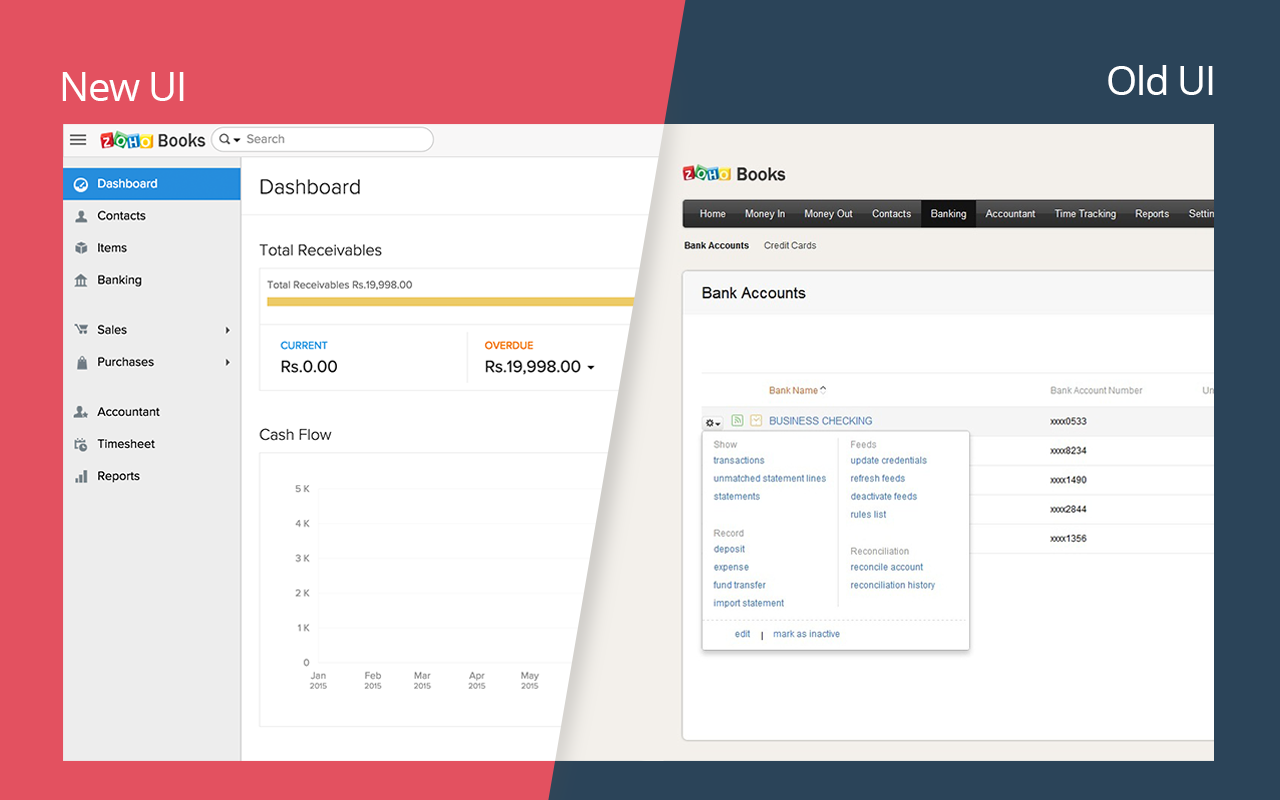}
\caption{Moving from tabbed navigation to left-hand side navigation \cite{jelumalai2015}}
\label{fig:navigation}
\end{figure}

It is not only sufficient that products simply look delightful but they have to work well without any issues as well. Any client of today's SaaS applications would have observed that most interfaces possess a left-hand side navigation as shown in Fig. \ref{fig:navigation} that is contrasted with the old pattern of utilizing tab navigation at the top of the screen \cite{jelumalai2015}.

For a great user experience design of SaaS, it is essential to make the key work areas to focus on simplicity of the design (easily accessible), and to avoid drawing attention to non-task related information (like unnecessary graphics, irrelevant images) which can distract users \cite{Knight2011}.

\subsubsection{Advantage}
This technique is recommended as it provides a feature to make it easy to the end users to resume interrupted tasks or recover from errors. Designing UI for SaaS applications are developed on top of designing efficient user goal oriented interactions \cite{Giurata2014}.

\subsubsection{Limitation}
In this method, conducting experiments with different users allows the development teams to consider every user's idea. The interface then developed tries to satisfy everybody but manages to satisfy nobody. This will lead to simply build it and correct it after user testing which is very costly \cite{Dubberly2001}.

\subsection{Advanced Responsive Design Technique}\label{exsol:rd}
\begin{figure}
\centering
\includegraphics[width=0.3\textwidth]{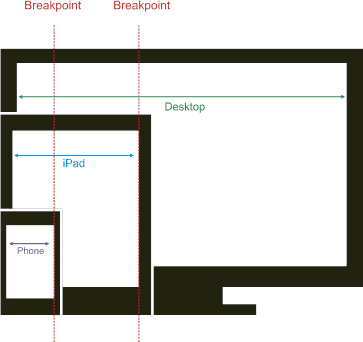}
\caption{Design considerations for responsive UI \cite{Varma2014}}
\label{fig:responsive}
\end{figure}

As SaaS service model grows more popular, UX becomes increasingly important. Websites, which are used to be responsive as shown in Fig. \ref{fig:responsive}, witness significant slowing down  because they attract traffic from unprecedented large number of users \cite{Varma2014}. This will ultimately result in losing customers because if an average online user does not get what he wants in less than 5 seconds, he would probably leave \cite{Mandryk2006}.

\subsubsection{Existing Solution}
In SaaS applications, this challenge could be rectified by adapting conditional loading which loads only what user's need as per their requirements which was assumed to be the requirements of the primary user. It follows thumb rule which first loads actual content, then enhancements and finally loads whatever leftovers \cite{Keith2011}. In mobile application, only the necessary elements have to be displayed as mobile friendly-designed websites only present a subset of information due to the smaller screen size \cite{Avery2015}.

\subsubsection{Advantage}
Conditional loading makes use of automation tools for scaling and caching various images. Any kind of changes to the site in the future will make it simpler and faster. However, it also depends on the user's internet connection as high user experience is achieved with better internet connection \cite{Keith2011}.

\subsubsection{Limitation}
According to \cite{Avery2015}, the bitter truth is that it does not apply to all interfaces and applications. Responsive web design does not identify the features provided by mobile devices and smart phones which is a threat for designers.

\subsection{User Experience Design Technique}\label{exsol:ux}
Different users from various domains and workplaces use SaaS applications. The user experience design challenge provides numerous possibilities to polish the interface design and to rectify related issues that can contribute to a positive user experience \cite{Garrett2010}.

\subsubsection{Existing Solution}
\begin{figure}
\centering
\includegraphics[width=0.4\textwidth]{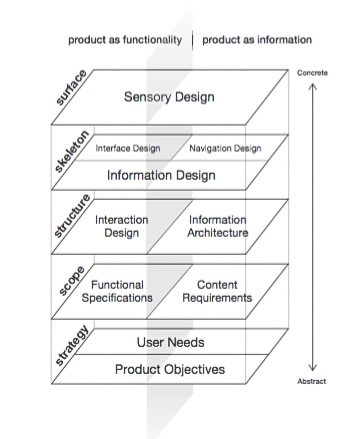}
\caption{The five planes of user experience \cite{Garrett2010}}
\label{fig:ux}
\end{figure}

Reference \cite{Garrett2010} presented a conceptual model to overcome the user experience design challenge in SaaS. Fig. \ref{fig:ux} shows that the model is splitting into two half based on the user's tasks. The left-side features are specific to functional aspects and on the right side there are features specific to information.

In the strategy plane, end user's needs and goals are the key design drivers that determine the whole strategy. In scope plane, the functional specifications and content requirements are well described. The structure plane describes the interface behavior through interaction design and information architecture. The information design, navigation design and interface design are designed by considering and rectifying the issues in the below level. This enables the end users to easily understand, navigate and interact with the application or product. The final surface plane is concerned about the issues related to visual appeal. This step by step consideration allows to rectify challenges in user experience design and provides a good solution to attain positive user experience \cite{Garrett2010}.

\subsubsection{Advantage}
The main strategy behind the model is to simplify the user experience into five interdependent planes which helps to design phase decisions and improves the design of overall user experience \cite{Garrett2010}.

\subsubsection{Limitation}
The UX approach starts from bottom to top and each plane is completely dependent on the evaluations made by planes below it. So, if one stage is designed incorrectly it will effect the entire five planes \cite{Garrett2010}.

\section{Conclusion and Future Work}
In SaaS applications, to overcome the user experience challenges it is very important to understand how the end users experience the new technology innovations. It is difficult to impress everyone having different behaviors and have one whole design/system that will appease/satisfy every user. This paper walks through a few key existing solutions to deal with UX challenges in SaaS application.

My next work will be conducting large scale online surveys sent out to large group of cloud computing experts in China, United States and Australia to validate identified UX challenges in SaaS applications. Future research directions include utilizing assertions and runtime monitoring \cite{zheng2014physically, zheng2015braceassertion, zheng2016efficient} to capture unknown user interactions, investigating the association between security analysis \cite{pan2017cyber, zheng2017security} and user interactions, and exploring the impact of increasingly popular Microservice adoption in SaaS \cite{yu2018survey, zheng2017bigvm, zheng2018smartvm, zheng2017smartvm} on user experience.

\bibliographystyle{IEEEtran}
\bibliography{IEEEabrv,reference}

\end{document}